# Ensuring Data Integrity in Electronic Health Records: A Quality Health Care Implication

P. Vimalachandran[1], H. Wang[2], Y. Zhang[2], B. Heyward[3] and F. Whittaker[2]

[1] *Centre for Applied Informatics, College of Engineering and Science,*
*Victoria University, Melbourne, Australia*
*Pasupathy.Vimalachandran@live.vu.edu.au*

[2] *Centre for Applied Informatics, College of Engineering and Science,*
*Victoria University, Melbourne, Australia*
*{hua.wang, yanchun.zhang, frank.whittaker }@vu.edu.au*

[3] *Nexus Online, Adelaide, Australia*
*benhey37@gmail.com*

**Abstract**

An Electronic Health Record (EHR) system must enable efficient availability of meaningful, accurate and complete data to assist improved clinical administration through the development, implementation and optimisation of clinical pathways. Therefore data integrity is the driving force in EHR systems and is an essential aspect of service delivery at all levels. However, preserving data integrity in EHR systems has become a major problem because of its consequences in promoting high standards of patient care. In this paper, we review and address the impact of data integrity of the use of EHR system and its associated issues. We determine and analyse three phases of data integrity of an EHR system. Finally, we also present an appropriate method to preserve the integrity in EHR systems. To analyse and evaluate the data integrity, one of the major clinical systems in Australia is considered. This will demonstrate the impact on quality and safety of patient care.

**Categories and Subject Descriptions**
Data integrity, healthcare.

## 1. Introduction

Today, more than ever, organisations realise the importance of data quality [1] because of the introduction of increased reliance upon networked data. However one of the serious problems in depending on networked data is 'dirty data'. Dirty data may include incomplete, missing or inaccurate information.

The concern is particularly significant in health care, where dirty data represents the dark side of the great potential offered by the adoption of health related IT systems. First and foremost, dirty data can lead to medical errors, which can kill or cause long-term damage to the health of patients. Data should be an accurate representation of its source. It should be reliable. Data should have internal consistency. Data should adhere to rules based on the logic of the real world. The accuracy, internal quality, and reliability of data are frequently referred as *data integrity* [2]. This means the enforcement of data integrity ensures the quality of the data.

In EHR, data integrity entails the accuracy of the complete health record's documentation. It encompasses information governance, patient identification and validation of authorship and record amendments. Furthermore the quality of data contained in an EHR is dependent on accurate information at the point of capture – the data source. For example, the Table1 shows the potential EHR risks and how these risks impact data integrity in healthcare.

As Table 1 exemplifies, while a primary goal of EHR implementation is the reduction of medical errors, reports of new types of errors directly related to EHR implementation that can compromise quality of care and patient safety have emerged [5].

Inaccurate health information may adversely affect the quality of an individual's healthcare. Maintaining the integrity and completeness of health data is paramount because the computerisation of health information grows and the scope of organisational exchange of health information widens into Health Information Exchanges (HIEs). Patient identity integrity is the accuracy, quality, and completeness of demographic data attached to or associated with an individual patient. This includes not only the accuracy and quality of the data as it relates to the individual, as well as the correctness of the linking or matching of all existing records for that individual within and across information systems. The quality of healthcare across the continuum depends on the integrity, reliability, and accuracy of health information [6,27, 36, 37, 38, 39].

**Table 1**. The importance of data integrity

| Potential EHR risk | Impacting data integrity example |
|---|---|
| Software bugs may jumble, delete, modify data or deposit it in the wrong spot | A laboratory value may come back with an extra character inadvertently inserted. |
| Software default settings | A patient's treatment for cancer was delayed by several years because a setting in her physician's EHR system defaulted to an old normal Pap test result instead of the more recent abnormal results [3] |
| Software internal programming error | Calculations like pounds to kilograms or Celsius to Fahrenheit. |
| Transcription error | A baby died from a massive drug overdose as a result of a transcription error that occurred when a handwritten order was entered into the computer system. However this medical error could have been prevented if automated alerts had been activated [4] |
| Inconsistencies between data fields | A structured data field may indicate that one pill should be taken twice a day, while the free-text filed says to take two pills in the morning and one pill in the evening |
| Copying and pasting information | Copying and pasting the same note accidentally for several rows may result in the same medication or condition repeated unnecessarily. |
| Templates default values | Templates automatically fill in data elements based on other data entries before clinicians complete the actual data. |
| Clinical environment may contribute to the occurrence of clinical decision support system error. | User distraction might cause data entry errors or inattentiveness to the information being presented by the decision support system. |

### 1.1. Motivation

With the continued advancement of electronic health records (EHRs), there is increasing concern that a potential loss of documentation integrity could lead to compromised patient care, care coordination, and quality reporting and research as well as fraud and abuse.

Poor system usability including inappropriate EHR design gets in the way of face-to-face interaction with patients and health care providers are forced to spend more time documenting required health information for the EHR. Features such as pop-up reminders, cumbersome menus and poor user interfaces can make EHRs far more time consuming than paper charts [7]. Poor system interface problems also can lead to poor decisions. For example, a laboratory value may come back with an extra character inadvertently inserted [8].

Inappropriate use of EHR can also result in potential data integrity issues. For example copy and paste or cloning can lead to redundant and inaccurate information in EHRs. Using this feature can cause authorship integrity issues since documentation cannot be tracked to the original source [9, 33-35].

The software system vendors often add functionalities to assist with documentation capture such as templates, use of standard phrases and paragraphs and automatic object insertion to improve efficiency of data capture, timelines, legibility, consistency and completeness of documentation. However, when used inappropriately, without proper education and controls, these features can lead to inaccurate documentation and potentially result in medical errors or allegations of fraud. For example, clinical values brought in from other parts of the electronic record [10, 30-32].

Although EHR related errors and their actual and potential impact on the quality of care and patient safety have been documented for years, more research still needs to be done to measure the occurrence of these errors and determine the causes to implement solutions.

### 1.2. Our Contribution

In this paper, we review the literature of the data integrity in EHR systems (Section 2). In Section 3,

we determine and analyse interrelated phases of data integrity of the EHR systems and examine the impact of data integrity in each phase. We propose an appropriate method to preserve the data integrity to reduce the EHR related issues and potential risks in Section 4 and finally Section 5 concludes and suggests future research opportunities.

## 2. Related work

While much has been written about EHR-associated risks impacting information integrity and the subsequent actual and potential impacts on quality of care and safety over at least the past decade, little has been done to systematically measure and analyse these risks, identify the root causes, and universally implement strategies (such as system design modifications and adoption of usability principles) to reduce the risks. However, attention to the potential unintended consequences of electronic documentation is growing [11,27-29, 32, 39]. In addition to the risks to the quality and safety of patient care, apprehension about EHR-associated errors may be a barrier to EHR adoption and use [12, 20, 21].

There are also no clear standards for defining, measuring, or analysing EHR-related errors [13]. The need of identifying and analysing the EHR related risks are paramount. There will be a number of risks that adversely affect the EHR environment. A software flaw in an EHR system containing hundreds or thousands of medical records, such as a glitch that causes an inaccurate recording of patients' allergies or medications, could adversely affect a large number of patients [14, 30]. These EHR systems design flaws also can result from improper system use and poor system usability [15]. EHR system software vendors include copy/paste and templates functionalities in capturing documentation and these inappropriate methods lead to EHR related errors [16, 31, 33].

Inadequate user training, human errors, disruption of system use or use of the system in ways not intended by the system developer can also be resulted errors in the EHR system. Use of decision support systems may lead to errors of omission, whereby individuals miss important data because the system does not prompt them to notice the information, or errors of commission, whereby individuals do what the system tells or allows them to do, even when it contradicts their training and other available information [17, 37].

## 3. Data integrity in EHR systems

The integrity of the entire EHR is reliant on the integrity of each of the following three phases shown in Figure 1. The process of ensuring integrity to each phase would be different.

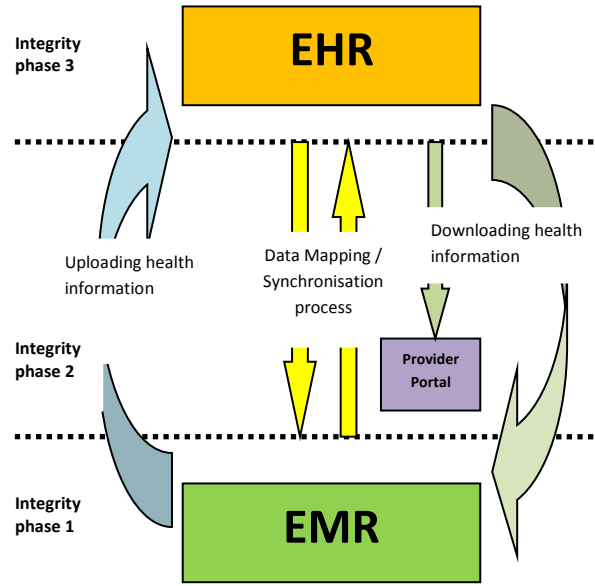

**Figure 1.** Integrity phases of EHR

### 3.1. Integrity phase 1: Ensuring data integrity in Electronic Medical Record (EMR) systems

Garbage in, garbage out. The data recorded into medical records that health care providers recorded must be meaningful and understandable for other health care providers when sharing health information. Otherwise it will not be useful for other healthcare providers who access those records when it is necessary. Providing the right information, at the right time for the right patient to deliver better health outcomes, must ensure the data integrity.

Ensuring data integrity in this phase must be carried out by clinical software systems. The data integrity must maintain the use of medical terminology or international medical coding system rather than free text usage in clinical related health information including medical condition and medical history. The coding system must be an option to choose from a pre developed item list to prevent spelling mistakes. The pre developed coding system must be linked to international medical approved standards dictionaries and updated on a regular basis, even daily. The clinical systems that health care providers use in Australia have included this facility. However this is deployed as an option and gives alternatives to add free texts to the clinical related information.

For example Figure 2 below, shows below how one of the major primary care provider clinical systems in Australia captures medical history item. There is always an option for clinicians to choose free text. The free text could lead to human errors or spelling mistakes and the impact might be very serious.

**Figure 2.** Medical history data capture form

A medical record in an EMR includes various medical related information of a patient. This information is recorded based on different types of data including patient's contributions, clinical check-up findings, pathology and radiology results and other measurements. The accuracy of health data of a patient depends on various inputs. The contribution from a patient is always based on the situation. The healthcare providers can only record what a patient communicates at the time of the consultation. In other words, a patient can intentionally hide a part of, or complete information from healthcare providers, they can even provide incorrect data. This integrity concern cannot be resolved unless the patient comes forward and provides the right information.

### 3.2. Integrity phase 2: Ensuring data integrity with linking right records

The communication of health information is a vital part of effective healthcare. The accurate identification of individuals is critical in all health communication. Mismatching of patients with their records and results is a documented problem for the health system and a clear link has been established between avoidable harm to patients and poor medical records management [18, 28, 29].

As mentioned above, identifying the right patient at the right time with the right information to upload or download is an essential part in this phase of integrity. To achieve this object, a standard number (e.g. index) system must be used. The system is intended to assist healthcare providers communicate health information with other providers accurately, for example, by providing a more reliable way of referencing patient information electronically.

Moreover, in EHR systems, the delivery of safe, effective and efficient healthcare relies on good communication and systems that share information, where subject of care can be reliably and consistently identified. In Australia, the Healthcare Identifier (HI) number system has been created and automatically allocated to all Australians who were enrolled with Medicare on 1st July 2010 [19].

HI is a unique 16 digit number used to identify patients that helps the healthcare providers to ensure their personal health information is linked with the right person. HIs are the building block for the Personally Controlled Electronic Health Record (PCEHR) system in Australia [20].
Medicare Australia uses the ISO7812 standard in order to create the HI number to every patient in Australia. The HI number contains, as shown in Figure 3 below, a single-digit Major Industry Identifier (MII), a six-digit Issuer Identifier Number (IIN), an Individual Account Identifier (IAI) number, and a single digit checksum based on the Luhn Algorithm. The MII forms the first part of the IIN.

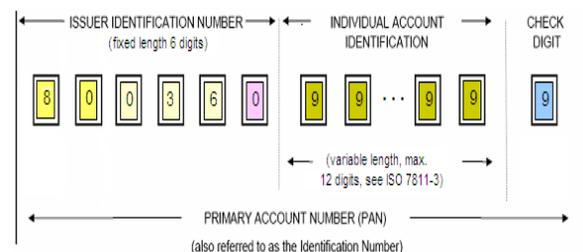

**Figure 3.** HI number system (Source:[21])

### 3.3. Integrity phase 3: Ensuring data integrity in EHR systems

Improved patient care, increased patient participation, improved care coordination, improved diagnostics and patient outcomes, practice efficiencies and cost savings are some of the major benefits of the EHRs [22]. However EHR over Internet will allow for the exposure of the records to theft and compromise. Personal details in the EHR including full name, date of birth, current address and Medicare number are valuable information for fraudsters to hack. The EHR also leads to deliver information to criminals which could be used to fraudulently obtain prescription drugs. This could have adverse implications for individuals, doctors and pharmacists whose e-health records are manipulated in order to facilitate criminal endeavours, where the audit trail will lead back to legitimate users who had access to these records, but who were in no way responsible for their fraudulent manipulation [23,34].

In 2012,Russian hackers held a Gold Coast medical centre to ransom after encrypting thousands of patient health records [24]. It was not a desktop that was compromised in this example. It was a server which is setup with much more security and rarely used for casual browsing and email. This means the hackers had to get through a firewall, anti-virus scanning software and then administrative credentials on the server.

Considering that recovering from a compromise is a non-trivial exercise, it is likely that these compromises persist for days or weeks, and some machines may remain compromised. Imagine if each of these computers had at least one user who had used it to access their PCEHR. That represents potentially millions of records compromised by online criminals.

That there will be a broad and extensive range of threats must be considered and managed to ensure the integrity of the PCEHR. These may cover the central infrastructure, including core server databases and data processing systems; intermediate data storage and processing systems used by healthcare professionals and service providers, and the data transport and communications layers, including protocols and channels used for end-to-end or server-to-server communications [23].

The integrity of the information of EHR needs to be trusted to use the system by the healthcare providers to ensure the success of the initiative. Some healthcare providers still have concerns on a few integrity issues. For example, the patients who registered with the PCEHR have got the ability to hide or completely delete (NOT to modify) the record that their healthcare providers uploaded for them. This control questions the integrity of the health information of the patient and how confident that the providers can rely on the system to provide the health care.

Lack of sense of shared accountability between system developers and users for product functionality of the EHR can lead to serious integrity issues. The department of health acknowledged to the Office of the Australian Information Commissioner (OAIC) that a technical change had introduced a glitch into the system potentially allowing a handful of healthcare providers to access PCEHR user's personal health notes without authorisation, for a short window of time [25].

## 4. Proposed model

The proposed model focuses on integrity phase three. Data encryption is commonly discussed and recommended for the EHR by many researches in the past.

We suggest pseudonymisation technique to protect the health information in the PCEHR. Adopting a pseudonym can preserve privacy. The sensitive data can be protected at the same time as allowing users access to less critical elements by means of a technique called pseudonymisation. This process handles sensitive data by substituting critical data elements with pseudonyms. Using this technique, the information cannot be accessed directly – only the specific elements of the data relevant to an enquiry are returned to the user. The result of the pseudonymisation process is that, although users can still search data for relationships, that user cannot capture all the value of the data outside the exact context of the interaction, and cannot amend it in an unauthorized manner at all. Copying pseudonymised data is similarly pointless, as the keys connecting the valuable links between the accessible pseudonym and the actual data itself are held elsewhere.

In the following example, Table 2 shows the actual data, however after the pseudonymisation process the sensitive data is hidden and the de-identified information is still useful and can be used for research purposes.

The hidden connective index data is stored in a secure destination or another PC where ordinary (or basic) users cannot access.

The difference between encryption and pseudonymisation is; encryption or password

**Table 2.** Basic pseudonymisation technique

| Healthcare Identifier | Medication | Date | Condition | Name |
|---|---|---|---|---|
| 8001567898761234 | Insulin | 01-10-2014 | CD | John Smith |
| 8008123456785000 | Dapotum | 05-10-2014 | MH | Jane Doe |
| 8001567898761234 | Thalitone | 10-10-2014 | CKD | John Smith |

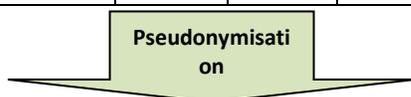
Pseudonymisation

| Healthcare Identifier ID | Medication | Date | Condition | Name |
|---|---|---|---|---|
| 0102 | Insulin | 01-10-2014 | CD | A12 |
| 452 | Dapotum | 05-10-2014 | MH | B02 |
| 2712 | Thalitone | 10-10-2014 | CKD | N17 |

**Table 3.** Hidden index data set

**Secure Data Store (stored in different secured destination or PC)**

Table A

| Healthcare Identifier | Healthcare Identifier Pseudonym |
|---|---|
| 8001567898761234 | 0102 |
| 8008123456785000 | 452 |
| 8001567898761234 | 2712 |

Table B

| Name | Name Pseudonym |
|---|---|
| John Smith | A12 |
| Jane Doe | B02 |
| John Smith | N17 |

permission exposes sensitive data and relationship. However, in pseudonymisation, the sensitive data is hidden and the relationships are exposed. The two key requirements for pseudonymisation are; data patterns must be maintained for linkage or analysis and personal data that will be shared, either internally or with a partner, must be hidden during the usage.

This will reduce risk exposure and mitigate any potential impact of internal and external security breaches. Pseudonymisation renders stolen data effectively useless for identity theft and other fraud. This facilitates secure outsourcing and off shoring by using de-identified data to identify accounts, process account documents and records accounts. The organizations can attain cost savings whilst significantly reducing the security concerns of using third party processors.

The health software system integrators, developers and systems administrators can use de-identified data for estimating eHealth projects that work with health sensitive data, designing and testing new systems that source health sensitive data from existing operations, and maintaining eHealth systems that manipulate sensitive data.

National health insurance, healthcare identifier and other health related number systems including Medicare effectively become sensitive through their long term usage. A good pseudonymisation solution can assign and maintain new pseudonyms for these sensitive identifiers as illustrated in Table 2 and 3 or even for different departments whilst coping with changes in these identifiers over time. However this is critical to business intelligence and other applications which deal frequently with historical data, in situations like mergers, acquisitions, and with time-related technical issues such as account number changes.

## 5. Conclusion and future suggestion

This paper presents three phases of integrity levels that require appropriate approaches to preserve data integrity for health in the usage of EHR systems. Pseudonymisation technique is proposed to preserve data integrity in EHR systems. With the introduction of pseudonymisation for the sensitive health information in EHR systems, indirect identifiable personal data is a long standing privacy to reduce the risks of identifiability. More research is needed on the industry-wide prevalence of each type of EHR risk and the impact on health record integrity, patient safety, and quality of care. Further research is also needed on the causes of EHR-related errors and effective strategies for preventing and correcting them.